\documentclass[aps,pre,twocolumn,showpacs,floatfix]{revtex4}
\usepackage{epsfig}

\newcommand     {\beq}[1]         { \begin{equation} #1 \end{equation} }
\newcommand     {\beqa}[1]        { \begin{eqnarray} #1 \end{eqnarray} }

\newcommand     {\EP}             { \varepsilon }
\newcommand     {\SI}             { \sigma }

\begin{document}

\title{Local load sharing fiber bundles with a
lower cutoff of strength disorder}

\author{Frank Raischel${}^1\footnote{Electronic 
address:raischel@ica1.uni-stuttgart.de}$, Ferenc Kun${}^{2}$, 
Hans J.\ Herrmann${}^{1,3}$}  
\affiliation{
\centerline{${}^1$ICP, University of Stuttgart, Pfaffenwaldring 27, D-70569
Stuttgart, Germany}
\centerline{${}^2$Department of Theoretical Physics, University of Debrecen, P.\
O.\ Box:5, H-4010 Debrecen, Hungary}  \\
\centerline{${}^3$IfB, ETH, Schafmattstr. 6, CH-8093 Z\"urich, Switzerland}
} 

\date{\today}
             
\begin{abstract}

We study the failure properties of fiber bundles with a finite lower
cutoff of the strength disorder varying the range of interaction
between the limiting cases of completely global and completely local
load sharing. Computer simulations revealed that at any range
of load redistribution there exists a critical cutoff strength where
the macroscopic response of the bundle becomes perfectly  
brittle, {\it i.e.} linearly elastic behavior is obtained up to
global failure, which occurs catastrophically after the breaking of
a small number of fibers. As an extension of recent mean field studies
[Phys.\ Rev.\ Lett.\ {\bf 95}, 125501 (2005)], we demonstrate that
approaching the critical cutoff, the size distribution of bursts of
breaking fibers shows a crossover to a universal power law form with
an exponent $3/2$ independent of the range of interaction. 
\end{abstract}
\maketitle

In engineering constructions solids are subjected to various types of
external loads and typically fail when the load exceeds a critical
value.  Monitoring stressed systems and forecasting an
imminent failure event is of enormous importance due to the possible  material
and human costs. In spite of the large amount of experimental and
theoretical efforts that has been undertaken over the past decades,
there is no comprehensive understanding of failure phenomena, which is
also reflected by the absence of reliable prediction
methods. 
Materials of low disorder typically fail in a ``one-crack''
way, where the main problem is to prevent crack initiation and
propagation. However, the failure of highly disordered materials
proceeds in bursts of local breaking events which can be 
recorded in form of acoustic signals. 
Experiments on a large variety of materials have revealed that 
in crackle noise spectra accompanying quasi-static fracture of
disordered materials, the amplitude and duration of signals and the
waiting time between them  are characterized by power law
distributions over a broad range 
\cite{garcimartin_prl79_1997,guarino_exp_acoust_epjb_1998,petri_prl73_1994}.
Quantitative changes of the burst activity when approaching the
critical load could be precursors of catastrophic failure and may
serve as the basis for forecasting techniques.

In the framework of discrete models of the fracture of disordered
materials, bursts can be identified as trails of correlated breakings
of the microscopic constituents of the model. 
Fiber bundle models consist of a parallel
bundle of fibers with identical linearly elastic behavior and
randomly distributed breaking thresholds
\cite{raul_burst_contdam,zapperi_prl78_1997,sornette_prl_78_2140}.  
Under an increasing external load, each fiber breaking is followed by
a load redistribution over the remaining intact fibers, which may
trigger avalanches of 
correlated breaking events analogous to crackling noise in experiments. 
Assuming global load sharing (GLS), it has been shown that the
distribution of avalanche sizes has a universal power law behavior
with an exponent $5/2$ \cite{kloster_pre_1997}.
In the other extreme of local load sharing (LLS), redistributing the
load solely over  the closest neighborhood of fibers, the avalanche
distribution appears also to be a power law but with a higher exponent
$\approx 9/2$ \cite{kloster_pre_1997,hansen_distburst_local_1994}. 
For global load sharing it has recently
been pointed out that the distribution of burst sizes significantly
changes if the weak fibers are removed from the bundle: if
the strength distribution of fibers has a finite lower
cutoff, or analogously, if the recording of avalanches starts after
the breaking of the weak elements, the burst size distribution is
found to show a crossover to another power law with a significantly lower
exponent $3/2$ \cite{hansen_crossover_prl,hansen_lower_cutoff_2005}. 
The effective range of interaction in real materials 
may have large variations \cite{raul_varint_2002}, therefore, in order
to use the crossover effect of burst sizes in forecasting of imminent
failure, its robustness with respect to the range of interaction has to
be explored. 

In the present paper we extend recent mean field studies of the effect
of the lower cutoff of fiber strength on the failure process of fiber
bundle models \cite{hansen_crossover_prl,hansen_lower_cutoff_2005}
by continuously varying the range of load sharing between the 
limiting cases of completely global load sharing and the very
localized one \cite{raul_varint_2002}. We show that at any
range of interaction there exists a critical value of the cutoff
strength above which the global response of the bundle becomes perfectly
brittle. We demonstrate that the 
crossover of the avalanche size distribution to a power law of an
exponent $3/2$, when approaching the critical cutoff strength, is
independent of the range of interaction. Our results support the usage of 
the crossover phenomenon of burst sizes in forecasting techniques of
imminent failure. 

We consider a parallel bundle of fibers organized on a square
lattice of size $L\times L$. The fibers are assumed to have linearly
elastic behavior with identical Young modulus $E$ up to a randomly
distributed breaking threshold. For simplicity, the failure thresholds
$\sigma_{th}$ are assumed to have a uniform distribution between a
lower cutoff strength $\sigma_{L}$ and one with the probability density
function $p(\sigma_{th})$ 
\begin{eqnarray}
\label{eq:disorder}
 p(\sigma_{th}) = \left\{ \begin{array}{rlcl}
	& \frac{1}{1-\sigma_{L}}, &\mbox{for}& \SI_{L} \leq \sigma_{th} \leq 1 \\ [3mm]
        & 0,                         &\mbox{otherwise.}& 
                          \end{array} \right. 
\end{eqnarray}
\begin{figure}%[!h]
\begin{center}
\epsfig{bbllx=12,bblly=8,bburx=700,bbury=312,file=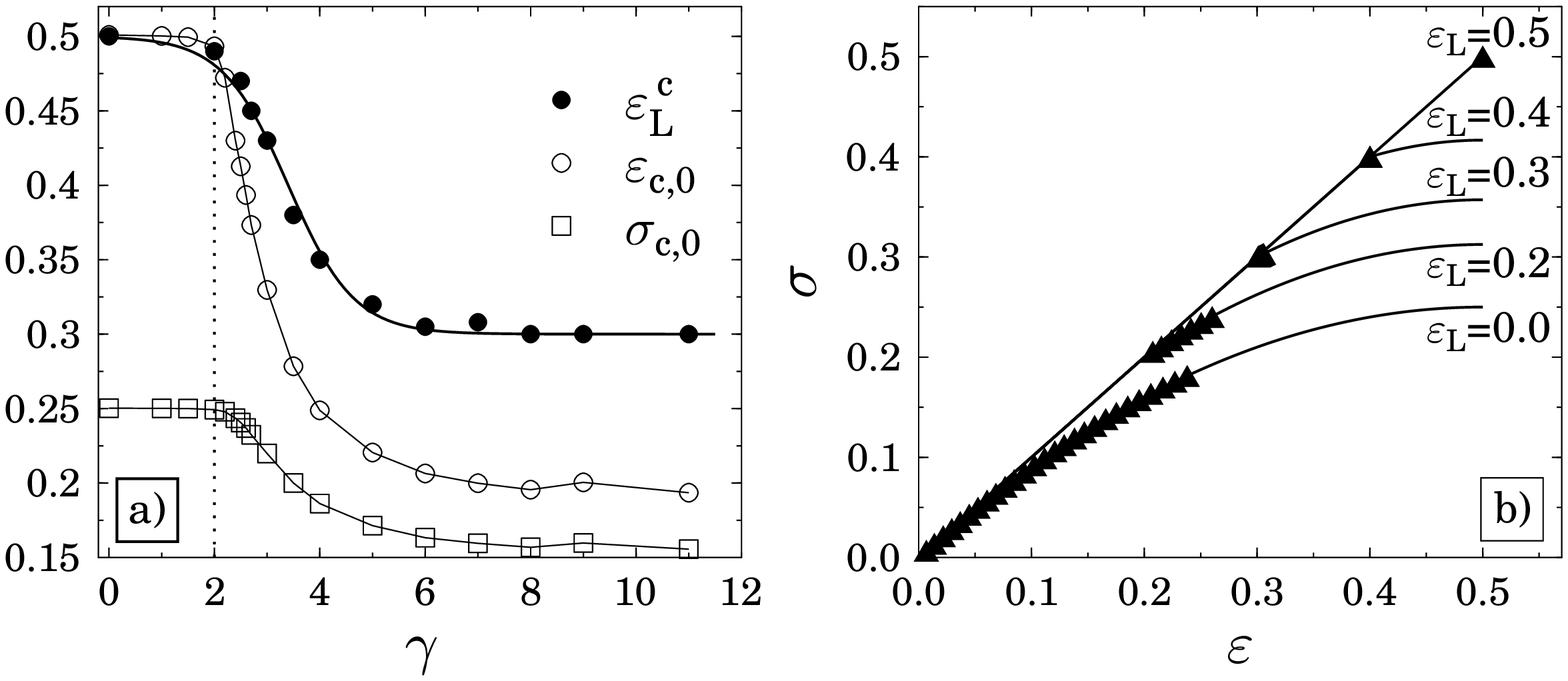,
  width=8.5cm}
   \caption{$a)$ Failure stress $\SI_c$ and strain $\EP_c$ of the fiber
bundle model at zero cutoff $\EP_L=0$ compared to the critical cutoff
$\EP_L^c$ as a function of $\gamma$. $b)$ Constitutive curves
for GLS $\gamma=0$ (lines) compared to the case of
$\gamma = 5$ (triangles) at different values of $\EP_L$.
 }
\label{fig:fig4}
  \end{center}
\end{figure}
Under an increasing external load 
the fibers break when the load on them exceeds the local threshold
value $\SI_{th}^i$, where $i=1, \ldots , N$ and $N=L^2$ denotes the
number of fibers. Due to the linearly elastic behavior, the failure
thresholds $\SI_{th}^i$ can also be expressed in terms of deformation
$\EP_{th}^i=\SI_{th}^i/E$ with the cutoff strength $\EP_L=\SI_L/E$.
After a failure event, the remaining intact fibers
have to take over the load of the failed one. In order to give a
realistic description of the load redistribution in FBMs,
we recently introduced a load transfer function of the form
\begin{eqnarray}
\sigma_{add} = \frac{1}{Z}r_{ij}^{-\gamma},
\label{eq:transfer}
\end{eqnarray}
where $\sigma_{add}$ denotes the additional load fiber $i$ receives
after the breaking of fiber $j$ \cite{raul_varint_2002}. The load
increment $\sigma_{add}$ decreases as a power law of the distance
$r_{ij}$ from the failed fiber, where the exponent 
$\gamma$ is considered to be a free parameter of the model. 
The exponent $\gamma$ can take any values between 0 and $\infty$
controlling the effective range of load redistribution between the
limiting cases of completely global $\gamma = 0$ and completely
localized load redistribution $\gamma \to \infty$ \cite{raul_varint_2002}.

Under perfectly global load sharing $\gamma = 0$ the macroscopic
constitutive equation of the system can be cast in a simple analytic
form  
\begin{eqnarray}
\sigma(\EP) = \left\{ \begin{array}{rlcl}
              & E\EP \ \ \ &\mbox{for}& \ \ \ E\EP \leq \sigma_{L} \\ [3mm]
              & E\EP\frac{1-E\EP}{1-E\EP_L}, \ \  &\mbox{for}& \ \
\SI_{L} \leq E\EP \leq 1,
              \end{array} \right.
\label{eq:constit}
\end{eqnarray}
where in the following, the value of the Young modulus of fibers will be set
to unity $E=1$. The constitutive behavior Eq.\ (\ref{eq:constit}) of
the bundle is perfectly linear up to the deformation 
$\EP_L$ since no fibers break in this regime (see also Fig.\
\ref{fig:fig4}$b$).  
Due to the breaking
of fibers above $\EP_L$, the constitutive curve $\SI(\EP)$ becomes
non-linear and develops a maximum whose value $\SI_c^{GLS}$
and position $\EP_c^{GLS}$ define the failure stress and strain of the
bundle, respectively. It follows from Eq.\ (\ref{eq:constit}) that the
critical strain is constant $\EP_c^{GLS}=1/2$ and does not depend on the
cutoff strength $\EP_{L}$, while $\SI_c^{GLS}$ increases due to the missing
weak fibers 
\beqa{
\sigma_c^{GLS} = \frac{1}{4(1-\EP_L)}. 
\label{eq:crit_sigma_gls}
}
Increasing the external load quasi-statically, the breaking fibers
trigger avalanches of failure events which either stop after a finite
fraction of fibers failed, or become unstable and destroy the entire
system. As a consequence, the cutoff 
strength $\EP_{L}$ can take meaningful values in the interval $0\leq 
\EP_{L} \leq \EP_c^{GLS}$, since for $\EP_{L} \geq \EP_c^{GLS}$ the
breaking of the first weakest fiber results in an immediate
catastrophic failure of the bundle.

\begin{figure}%[!h]
\begin{center}
\epsfig{bbllx=135,bblly=185,bburx=610,bbury=610,file=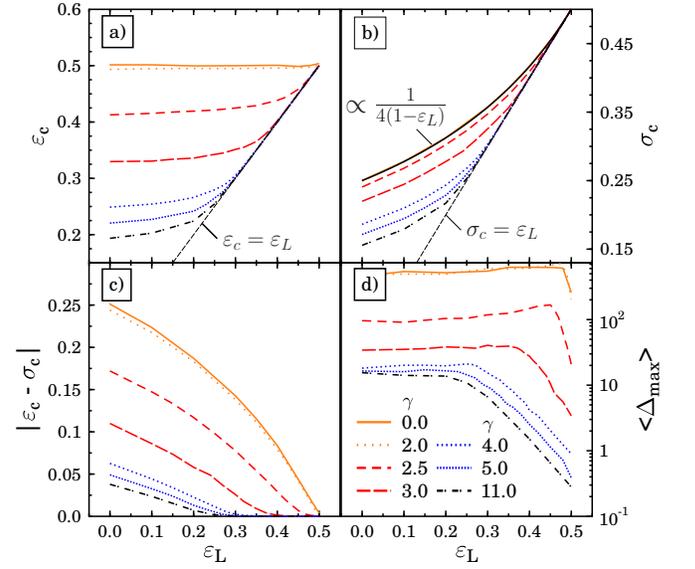,
  width=8.5cm}
   \caption{ Characteristic quantities of the failure process of FBM
varying the effective range of load sharing $\gamma$ and the cutoff
value of failure strength $\EP_L$: $a)$ critical deformation $\EP_c$,
$b)$ critical stress $\SI_c$, $c)$  $| \EP_c-\SI_c|$, $d)$ the average size of the largest avalanche
$\left<\Delta_{max}\right>$ as a function of $\EP_L$ for various
values of $\gamma$.}
\label{fig:fig1}
  \end{center}
\end{figure}
We explore the effect of the finite cutoff strength
$\EP_L$ on the failure process of FBMs with short ranged load sharing 
by means of computer simulations, redistributing the load
of broken fibers according to the load transfer function Eq.\
(\ref{eq:transfer}). Stress controlled simulations were carried out
on a square lattice of size $L=257$ with periodic boundary conditions
varying the  cutoff strength $\EP_L$ of the disorder distribution Eq.\
(\ref{eq:disorder}) in the interval $[0, 0.5]$ at several different
values of the effective range of interaction $\gamma$ between 0 and 11.
To characterize the failure process of the bundle at the macro and
micro level, we determined the critical stress $\SI_c$ and strain
$\EP_c$, the distribution $D$ 
of avalanche sizes $\Delta$, the average avalanche size $\left< \Delta
\right>$, and the average value of the largest avalanche $\left<
\Delta_{max}\right>$. 
For clarity, we first characterize the behavior of the system in the
specific case of zero cutoff $\EP_L=0$ by studying the critical stress
$\SI_c$ and strain $\EP_c$ of the bundle as a function of $\gamma$,
see Fig.\ \ref{fig:fig4}$a$. Based on the numerical results, 
three regimes of the failure of FBM can be distinguished in Fig.\ 
\ref{fig:fig4}$a$ depending on the range of load sharing: for $\gamma 
\leq 2$ the range of interaction is infinite in the two dimensional
embedding space, hence both $\SI_c$
and $\EP_c$ take their GLS values  $\SI_c=0.25$ and
$\EP_c=0.5$ independent of $\gamma$ (see Eq.\
(\ref{eq:crit_sigma_gls}) at $\EP_L=0$). Increasing the value 
of $\gamma \geq 2$ the effective range of interaction gradually decreases
which lowers  the macroscopic strength $\EP_c$ and $\SI_c$ of the
bundle. In the limiting case of 
$\gamma \to \infty$ the model recovers the very localized load
sharing, where $\EP_c$ and $\SI_c$ take again constant
values. According to the numerical results, the perfectly localized
limit is practically 
reached for $\gamma \geq 6$, so that in the interval  $2 \le \gamma \le
6$ a transition occurs between the completely global and completely
local behavior \cite{raul_varint_2002}. 
Fig.\ \ref{fig:fig4}$b$ demonstrates that for $\gamma \leq 2$ ({\it i.e.} GLS) the
macroscopic failure of the bundle is preceded by a strong
non-linearity of the constitutive curve $\SI(\EP)$. At any wider range of
load sharing, $\gamma>2$, the $\SI(\EP)$ curves follow the GLS solution Eq.\
(\ref{eq:constit}), but with lower strength values which implies a more
brittle macroscopic response for short ranged interactions. 

Varying the cutoff strength $\EP_L$ at different values of $\gamma$, 
it can be seen in Figs.\ \ref{fig:fig1}$a,b)$ that in the long range regime
$\gamma \leq 2$ both $\EP_c$ and $\SI_c$ agree well with the analytic
predictions Eq.\ (\ref{eq:crit_sigma_gls}), {\it i.e.} $\EP_c=1/2$
is constant while $\SI_c$ increases with increasing cutoff $\EP_L$. 
When the load sharing becomes short ranged $\gamma > 2$, the increasing
macroscopic brittleness has the consequence that the curves of
$\EP_c(\EP_L)$ and 
$\SI_c(\EP_L)$ shift downwards as $\gamma$ increases and tend to a
limit curve when the interaction becomes completely localized for $\gamma
\geq 6$. It is interesting to note that for short range interaction of
fibers $\gamma >2$, not only the failure stress $\SI_c$ but also the
failure strain $\EP_c$ is an increasing function of $\EP_L$.
It is important to emphasize that at each $\gamma$ there exists a
critical value of the cutoff strength $\EP_L^c < \EP_c^{GLS}$ where the
failure stress $\SI_c$ and strain $\EP_c$ of the system becomes equal
to the cutoff strength, {\it i.e.} at $\EP_L^c$ holds
$\EP_c(\EP_L^c)=\SI_c(\EP_L^c)=\EP_L^c$. At this point the macroscopic
response of the bundle becomes perfectly brittle, {\it i.e.} under
gradual loading of the system the macroscopic constitutive behavior
is linear up to $\SI_c$, where the breaking of the weakest fiber gives
rise to the collapse of the entire system (Fig.\
\ref{fig:fig4}$b$). This transition is better 
illustrated by Fig.\ \ref{fig:fig1}$c$ where the difference
$\delta = |\EP_c(\EP_L)-\SI_c(\EP_L)|$ is plotted
versus $\EP_L$. It can be observed that $\delta$ monotonically
decreases and becomes practically zero at $\EP_L^c$ of the given
$\gamma$. Since the absence of weak fibers gives rise to a higher
macroscopic strength, the value of $\EP_L^c$ is larger than the
strength of the bundle $\EP_c$ and $\SI_c$ at zero cutoff (see Fig.\
\ref{fig:fig4}$a$). 

\begin{figure}%[!h]
\begin{center}
\epsfig{bbllx=0,bblly=0,bburx=450,bbury=450,file=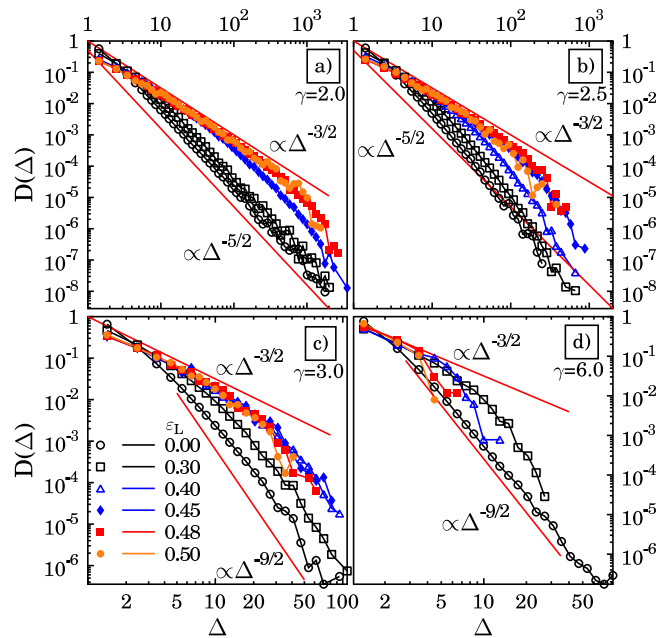,
  width=8.5cm}
   \caption{ Distribution of burst sizes $D(\Delta)$ varying  $\EP_L$ at different values of $\gamma$: $a)$ 2.0,  $b)$ 2.5,  $c)$
3.0,  $d)$ 6.0. Crossover behavior of $D(\Delta)$ can be observed as
$\EP_L$ approaches the critical cutoff value $\EP_L^c(\gamma)$.  
 }
\label{fig:fig2}
  \end{center}
\end{figure}
On the micro level, the failure process is characterized by the bursts
of fiber breakings, which also show an interesting behavior when the
range of interaction $\gamma$ and the lower cutoff $\EP_L$ are varied. In
the GLS regime $\gamma < 2$, our computer simulations perfectly
recover the analytical and numerical results of Refs.\
\cite{hansen_crossover_prl,hansen_lower_cutoff_2005} (see Fig.\
\ref{fig:fig2}$a$): for $\EP_L=0$ the 
size distribution of bursts $D(\Delta)$ follows a power law
\beq{
D(\Delta) \sim \Delta^{-\alpha},
\label{eq:burst_gls}
}
with an exponent $\alpha = 5/2$. Increasing the value of the cutoff
$\EP_L$, for small avalanches a crossover occurs to a power law of a
lower exponent $\alpha = 3/2$, while for large avalanches the original
power law with $\alpha=5/2$ is retained. The crossover avalanche size
$\Delta_c$ increases with increasing $\EP_L$ and the distribution
becomes a single power law of exponent $\alpha = 3/2$ as $\EP_L$
approaches $\EP_c^{GLS}$. The crossover to a lower value of the
exponent indicates that due to the missing weak fibers the fraction of
small avalanches decreases compared to the larger ones.  
\begin{figure}%[!h]
\begin{center}
\epsfig{bbllx=40,bblly=40,bburx=360,bbury=310,file=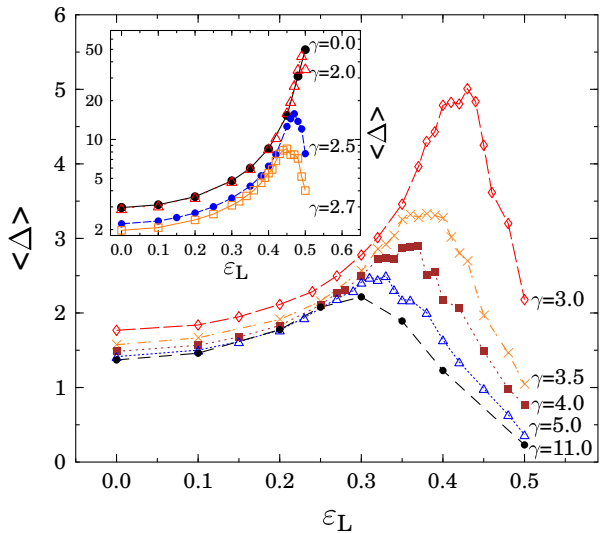,
  width=8.0cm}
   \caption{ The average burst size $\left< \Delta \right>$ as a
function of $\EP_L$ for different values of
$\gamma$. In the short range regime $\gamma > 2$, the average
avalanche size  $\left< \Delta \right>$ has a clear maximum which
coincides with the critical cutoff strength $\EP_L^c$. 
 }
\label{fig:fig3}
  \end{center}
\end{figure}
This argument is further supported by Fig.\ \ref{fig:fig1}$d$ and
Fig.\ \ref{fig:fig3} which demonstrate that for $\gamma \leq 2$ both
the average size of
the largest avalanche $\left< \Delta_{max}\right>$ and the average
avalanche size $\left< \Delta \right>$ are monotonically increasing
functions of the cutoff $\EP_L$. However, when the
load sharing gets short ranged $\gamma > 2$, both $\left<
\Delta_{max}\right>$ and $\left< \Delta \right>$ have a maximum at the
critical cutoff strength. Note that the non-vanishing avalanche size
above $\EP_L^c$ arises due to the strength fluctuations of the finite
bundle so that above $\EP_L^c$ the bundle may survive a small number of
avalanches instead of collapsing after the breaking of the weakest
fiber. It is interesting to note that 
contrary to GLS, in the transition regime $2 < \gamma < 6$, 
the avalanche size distribution does not show a power law
behavior for small cutoffs $\EP_L \approx 0$, however, when
$\EP_L$ approaches the critical value $\EP_L^c(\gamma)$, the
distribution of burst sizes $D(\Delta)$ tends again to a power law of 
an exponent $\alpha=3/2$ (Fig.\ \ref{fig:fig2}$b,c$). For very
localized interactions $\gamma > 6$ 
an apparent power law of $D(\Delta)$ is restored for
$\EP_L\approx 0$ with a relatively high exponent $\alpha \approx 9/2$,
in agreement with Ref.\ \cite{hansen_distburst_local_1994} (Fig.\
\ref{fig:fig2}$d$). The main outcome of computer simulations is that
the crossover behavior of $D(\Delta)$ to the universal power law
$D(\Delta) \sim \Delta^{-3/2}$ prevails at any value of the range of
interaction $\gamma$ for the limiting case of $\EP_L \to
\EP_L^c(\gamma)$, independently of the original form of $D(\Delta)$ at
zero cutoff $\EP_L=0$ (see Fig.\ \ref{fig:fig2}). 
In spite of the relatively large system size $L$, for short range
interaction of fibers and $\EP_L \to \EP_L^c(\gamma)$ the statistics
of avalanche sizes is rather poor for large avalanches which hinders
us to make a definite conclusion on 
the shape of $D(\Delta)$ in this $\Delta$ regime.

An interesting experimental realization of the crossover for crackling noise was very recently found
in the magnitude distribution of earthquakes in Japan
\cite{kawamura_crossover_2006}. Analyzing the local magnitude
distribution of earthquakes preceding main shocks,
a significant decrease of the Gutenberg-Richter exponent was obtained
when the lower bound of the time window of the analysis is shifted
towards the catastrophic event
\cite{kawamura_crossover_2006}. Fracture of ferromagnetic materials is
accompanied by changes of the magnetic flux, which can be recorded as magnetic noise and provides
information on the dynamics of crack propagation
\cite{kun_magnetnoise_prl_2004}. The amplitude, area and energy of
magnetic emission signals have recently been found to have power law
distributions with exponents depending on the type of fracture, {\it
i.e.} ductile failure where stable crack propagation occurs in a large
number of elementary steps is characterized by significantly higher
exponents than brittle failure, where the crack propagates in an
unstable catastrophic manner breaking the specimen in a few large
jumps \cite{kun_magnetnoise_prl_2004}. Our numerical results suggest
that the reduction of non-linearity of the macroscopic response of
materials preceding global failure when going from ductile and
quasi-brittle to brittle fracture is responsible for the lowering of the
crackling noise exponents on the micro-level.

In summary, we carried out computer simulations of the failure process
of a bundle of fibers with a finite cutoff of the fibers'
strength, continuously varying the range of interaction between the
limiting cases of global and local load sharing. 
We showed that increasing the cutoff
strength $\EP_L$ the macroscopic response of the fiber bundle becomes
perfectly brittle when $\EP_L$ approaches a critical value
$\EP_L^c(\gamma)$, depending on the range of interaction $\gamma$. 
Our numerical results demonstrate the robustness of the crossover of
the avalanche size distribution $D(\Delta)$ to a universal power law
of exponent $3/2$, irrespective of the range of interaction between the
material elements.

\vspace*{-0.6cm}
\begin{acknowledgments}
This work was supported by the Collaborative Research Center SFB381. F.\ Kun
acknowledges financial support of the Research Contracts 
NKFP-3A/043/04, OTKA M041537, T049209.
\end{acknowledgments}
\vspace*{-0.3cm}

\bibliography{../statphys_fracture.bib}
\end{document}